\input harvmac
\input epsf

\def\figin{\epsfcheck\figin}\def\figins{\epsfcheck\figins}
\def\epsfcheck{\ifx\epsfbox\UnDeFiNeD
\message{(NO epsf.tex, FIGURES WILL BE IGNORED)}
\gdef\figin##1{\vskip2in}\gdef\figins##1{\hskip.5in}% blank space instead
\else\message{(FIGURES WILL BE INCLUDED)}%
\gdef\figin##1{##1}\gdef\figins##1{##1}\fi}
\def\DefWarn#1{}
\def\figinsert{\goodbreak\topinsert}
\def\ifig#1#2#3#4{\DefWarn#1\xdef#1{fig.~\the\figno}
\writedef{#1\leftbracket fig.\noexpand~\the\figno}%
\figinsert\figin{\centerline{\epsfxsize=#3mm \epsfbox{#2}}}
\bigskip\medskip\centerline{\vbox{\baselineskip12pt
\advance\hsize by -1truein\noindent\footnotefont{\sl Fig.~\the\figno:}\sl\ #4}}
\bigskip\endinsert\noindent\global\advance\figno by1}

\def\a{\alpha}
\def\b{\beta}

\def\th{\theta}

\def\F{\Phi}

\def\Tr{{\rm Tr}}
\def\hf{{1\over 2}}

\def\cF{{\cal F}}
\def\cN{{\cal N}}
\def\cW{{\cal W}}

\def\({\bigl(}
\def\){\bigr)}
\def\<{\langle}
\def\>{\rangle}

\def\Im{{\rm Im}\;}
\def\mbar{\overline{m}}
\def\Fbar{\overline{\F}}

\def\Phib{\Fbar}
\def\Db{\overline{D}}

\def\ad{{\dot{\alpha}}}

\def\Del{\nabla}
\def\Delb{\overline{\nabla}}

\def\frac#1#2{{#1 \over #2}}
\def\em{\it}

\def\sqr#1#2{{%\vcenter
{\vbox{\hrule height.#2pt
\hbox{\vrule width.#2pt height#1pt \kern#1pt
\vrule width.#2pt} \hrule height.#2pt}}}}

\def\Box{{\,\sqr66\,}}

\lref\thooft{
G.~'t Hooft, ``A Planar Diagram Theory For Strong Interactions,''
Nucl.\ Phys.\ B {\bf 72}, 461 (1974).}

\lref\mm{
P.~Ginsparg and G.~W.~Moore,
``Lectures On 2-D Gravity And 2-D String Theory,''
arXiv:hep-th/9304011.}

\lref\mmm{
P.~Di Francesco, P.~Ginsparg and J.~Zinn-Justin,
``2-D Gravity and random matrices,''
Phys.\ Rept.\  {\bf 254}, 1 (1995)
[arXiv:hep-th/9306153]}

\lref\kontsevich{M.~Kontsevich,
``Intersection Theory On The Moduli Space Of Curves And The Matrix
Airy Function,'' Commun.\ Math.\ Phys.\ {\bf 147}, 1 (1992).}

\lref\wittentop{E.~Witten,
``On The Structure Of The Topological Phase Of Two-Dimensional
Gravity,'' Nucl.\ Phys.\ B {\bf 340}, 281 (1990)}

\lref\bfss{T.~Banks, W.~Fischler, S.~H.~Shenker and L.~Susskind,
``M theory as a matrix model: A conjecture,'' Phys.\ Rev.\ D {\bf 55},
5112 (1997) [arXiv:hep-th/9610043].}

\lref\adscft{
O.~Aharony, S.~S.~Gubser, J.~M.~Maldacena, H.~Ooguri and Y.~Oz,
``Large $N$ field theories, string theory and gravity,'' Phys.\ Rept.\
{\bf 323}, 183 (2000) [arXiv:hep-th/9905111].  }

\lref\ghv{D. Ghoshal and C. Vafa, ``$c=1$ string as the topological theory
of the conifold,'' Nucl.\ Phys.\ B {\bf 453}, 121 (1995)
[arXiv:hep-th/9506122].}

\lref\klmkv{S. Kachru, A. Klemm, W. Lerche, P. Mayr, C. Vafa,
``Nonperturbative Results on the Point Particle Limit of $N=2$
Heterotic String Compactifications,'' Nucl.\ Phys.\ B {\bf 459}, 537
(1996) [arXiv:hep-th/9508155].}

\lref\klw{A. Klemm, W. Lerche, P. Mayr, C.Vafa, N. Warner,
``Self-Dual Strings and $N=2$ Supersymmetric Field Theory,''
 Nucl.\ Phys. \ B {\bf 477}, 746 (1996)
[arXiv:hep-th/9604034].}

\lref\kmv{S. Katz, P. Mayr, C. Vafa,
``Mirror symmetry and Exact Solution of 4D $N=2$ Gauge Theories I,''
Adv.\ Theor.\ Math.\ Phys. {\bf 1}, 53 (1998)
[arXiv:hep-th/9706110].}

\lref\gv{R.~Gopakumar and C.~Vafa,
``On the gauge theory/geometry correspondence,'' Adv.\ Theor.\ Math.\
Phys.\ {\bf 3}, 1415 (1999) [arXiv:hep-th/9811131].  }

\lref\edel{J.D. Edelstein, K. Oh and R. Tatar, ``Orientifold,
geometric transition and large $N$ duality for $SO/Sp$ gauge
theories,'' JHEP {\bf 0105}, 009 (2001) [arXiv:hep-th/0104037].}

\lref\dasg{K. Dasgupta, K. Oh and R. Tatar, ``Geometric transition, large
$N$ dualities and MQCD dynamics,'' Nucl. Phys.  B {\bf 610}, 331
(2001) [arXiv:hep-th/0105066]\semi -----, ``Open/closed string
dualities and Seiberg duality from geometric transitions in
M-theory,'' [arXiv:hep-th/0106040]\semi -----, ``Geometric transition
versus cascading solution,'' JHEP {\bf 0201}, 031 (2002)
[arXiv:hep-th/0110050].}

\lref\hv{K. Hori and C. Vafa,
``Mirror Symmetry,'' [arXiv:hep-th/0002222].}

\lref\hiv{K. Hori, A. Iqbal and C. Vafa,
``D-Branes And Mirror Symmetry,'' [arXiv:hep-th/0005247].}

\lref\vaug{C.~Vafa,
``Superstrings and topological strings at large $N$,''
J.\ Math.\ Phys.\  {\bf 42}, 2798 (2001)
[arXiv:hep-th/0008142].}

\lref\civ{
F.~Cachazo, K.~A.~Intriligator and C.~Vafa,
``A large $N$ duality via a geometric transition,''
Nucl.\ Phys.\ B {\bf 603}, 3 (2001)
[arXiv:hep-th/0103067].}

\lref\ckv{
F.~Cachazo, S.~Katz and C.~Vafa,
``Geometric transitions and $N = 1$ quiver theories,''
arXiv:hep-th/0108120.}

\lref\cfikv{
F.~Cachazo, B.~Fiol, K.~A.~Intriligator, S.~Katz and C.~Vafa, ``A
geometric unification of dualities,'' Nucl.\ Phys.\ B {\bf 628}, 3
(2002) [arXiv:hep-th/0110028].}

\lref\cv{
F.~Cachazo and C.~Vafa, ``$N=1$ and $N=2$ geometry from fluxes,''
arXiv:hep-th/0206017.}

\lref\ov{
H.~Ooguri and C.~Vafa, ``Worldsheet derivation of a large $N$ duality,''
arXiv:hep-th/0205297.}

\lref\av{
M.~Aganagic and C.~Vafa, ``$G_2$ manifolds, mirror symmetry, and
geometric engineering,'' arXiv:hep-th/0110171.}

\lref\digra{
D.~E.~Diaconescu, B.~Florea and A.~Grassi, ``Geometric transitions and
open string instantons,'' arXiv:hep-th/0205234.}

\lref\amv{
M.~Aganagic, M.~Marino and C.~Vafa, ``All loop topological string
amplitudes from Chern-Simons theory,'' arXiv:hep-th/0206164.}

\lref\dfg{
D.~E.~Diaconescu, B.~Florea and A.~Grassi, ``Geometric transitions,
del Pezzo surfaces and open string instantons,''
arXiv:hep-th/0206163.}

\lref\kkl{
S.~Kachru, S.~Katz, A.~E.~Lawrence and J.~McGreevy, ``Open string
instantons and superpotentials,'' Phys.\ Rev.\ D {\bf 62}, 026001
(2000) [arXiv:hep-th/9912151].}

\lref\bcov{
M.~Bershadsky, S.~Cecotti, H.~Ooguri and C.~Vafa, ``Kodaira-Spencer
theory of gravity and exact results for quantum string amplitudes,''
Commun.\ Math.\ Phys.\ {\bf 165}, 311 (1994) [arXiv:hep-th/9309140].
}

\lref\witcs{
E.~Witten, ``Chern-Simons gauge theory as a string theory,''
arXiv:hep-th/9207094.  }

\lref\naret{I. Antoniadis, E. Gava, K.S. Narain, T.R. Taylor,
``Topological Amplitudes in String Theory,'' Nucl.\ Phys.\ B\ {\bf
413}, 162 (1994) [arXiv:hep-th/9307158].}

\lref\witf{E. Witten,
``Solutions Of Four-Dimensional Field Theories Via M Theory,'' Nucl.\
Phys.\ B\ {\bf 500}, 3 (1997) [arXiv:hep-th/9703166].}

\lref\shenker{
S.~H.~Shenker, ``The Strength Of Nonperturbative Effects In String
Theory,'' in Proceedings Cargese 1990, {\it Random surfaces and
quantum gravity}, 191--200.  }

\lref\berwa{
M.~Bershadsky, W.~Lerche, D.~Nemeschansky and N.~P.~Warner, ``Extended
$N=2$ superconformal structure of gravity and W gravity coupled to
matter,'' Nucl.\ Phys.\ B {\bf 401}, 304 (1993)
[arXiv:hep-th/9211040]}

\lref\loopequation{
G.~Akemann, ``Higher genus correlators for the Hermitian matrix model
with multiple cuts,'' Nucl.\ Phys.\ B {\bf 482}, 403 (1996)
[arXiv:hep-th/9606004].  }

\lref\wiegmann{P.B. Wiegmann and A. Zabrodin, ``Conformal maps
and integrable hierarchies,'' arXiv:hep-th/9909147.}

\lref\kazakov{
S.~Y.~Alexandrov, V.~A.~Kazakov and I.~K.~Kostov, ``Time-dependent
backgrounds of 2D string theory,'' arXiv:hep-th/0205079.}

\lref\dj{
S.~R.~Das and A.~Jevicki, ``String Field Theory And Physical
Interpretation Of $D=1$ Strings,'' Mod.\ Phys.\ Lett.\ A {\bf 5}, 1639
(1990).}

\lref\givental{
A.B.~Givental, ``Gromov-Witten invariants and quantization of
 quadratic hamiltonians,'' arXiv:math.AG/0108100.}

\lref\op{
A.~Okounkov and R.~Pandharipande, ``Gromov-Witten theory, Hurwitz
theory, and completed cycle,'' arXiv:math.AG/0204305.}

\lref\dijk{ R.~Dijkgraaf, ``Intersection theory, integrable hierarchies and
topological field theory,'' in Cargese Summer School on {\it New
Symmetry Principles in Quantum Field Theory} 1991,
[arXiv:hep-th/9201003].}

\lref\gw{
D.~J.~Gross and E.~Witten, ``Possible Third Order Phase Transition In
The Large $N$ Lattice Gauge Theory,'' Phys.\ Rev.\ D {\bf 21}, 446
(1980).}

\lref\sw{
N.~Seiberg and E.~Witten, ``Electric-magnetic duality, monopole
condensation, and confinement in $N=2$ supersymmetric Yang-Mills
theory,'' Nucl.\ Phys.\ B {\bf 426}, 19 (1994) [Erratum-ibid.\ B {\bf
430}, 485 (1994)] [arXiv:hep-th/9407087].}

\lref\kkv{
S.~Katz, A.~Klemm and C.~Vafa, ``Geometric engineering of quantum
field theories,'' Nucl.\ Phys.\ B {\bf 497}, 173 (1997)
[arXiv:hep-th/9609239].}

\lref\taylor{
W.~I.~Taylor, ``D-brane field theory on compact spaces,'' Phys.\
Lett.\ B {\bf 394}, 283 (1997) [arXiv:hep-th/9611042].}

\lref\kmmms{
S.~Kharchev, A.~Marshakov, A.~Mironov, A.~Morozov and S.~Pakuliak,
``Conformal matrix models as an alternative to conventional
multimatrix models,'' Nucl.\ Phys.\ B {\bf 404}, 717 (1993)
[arXiv:hep-th/9208044].}

\lref\kostov{
I.~K.~Kostov, ``Gauge invariant matrix model for the A-D-E closed
strings,'' Phys.\ Lett.\ B {\bf 297}, 74 (1992)
[arXiv:hep-th/9208053].  }

\lref\ot{
K.~h.~Oh and R.~Tatar, ``Duality and confinement in $N=1$
supersymmetric theories from geometric transitions,''
arXiv:hep-th/0112040.}

\lref\ovknot{
H.~Ooguri and C.~Vafa, ``Knot invariants and topological strings,''
Nucl.\ Phys.\ B {\bf 577}, 419 (2000) [arXiv:hep-th/9912123].  }

\lref\dvv{
R.~Dijkgraaf, H.~Verlinde and E.~Verlinde, ``Loop Equations And
Virasoro Constraints In Nonperturbative 2-D Quantum Gravity,'' Nucl.\
Phys.\ B {\bf 348}, 435 (1991).}

\lref\kawai{
M.~Fukuma, H.~Kawai and R.~Nakayama, ``Continuum Schwinger-Dyson
Equations And Universal Structures In Two-Dimensional Quantum
Gravity,'' Int.\ J.\ Mod.\ Phys.\ A {\bf 6}, 1385 (1991).}

\lref\nekrasov{
N.~A.~Nekrasov, ``Seiberg-Witten prepotential from instanton
counting,'' arXiv:hep-th/0206161.}

\lref\ki{
I.~K.~Kostov, ``Bilinear functional equations in 2D quantum gravity,''
in Razlog 1995, {\it New trends in quantum field theory}, 77--90,
[arXiv:hep-th/9602117].}

\lref\kii{
I.~K.~Kostov, ``Conformal field theory techniques in random matrix
models,'' arXiv:hep-th/9907060.}

\lref\morozov{
A.~Morozov, ``Integrability And Matrix Models,'' Phys.\ Usp.\ {\bf
37}, 1 (1994) [arXiv:hep-th/9303139].}

\lref\fo{
H.~Fuji and Y.~Ookouchi, ``Confining phase superpotentials for SO/Sp
gauge theories via geometric transition,'' arXiv:hep-th/0205301.  }

\lref\marinor{M. Marino, ``Chern-Simons theory, matrix integrals, and
perturbative three-manifold invariants,'' [arXiv:hep-th/0207096].}

\lref\agnt{
I.~Antoniadis, E.~Gava, K.~S.~Narain and T.~R.~Taylor, ``Topological
amplitudes in string theory,'' Nucl.\ Phys.\ B {\bf 413}, 162 (1994)
[arXiv:hep-th/9307158].}

\lref\vw{
C.~Vafa and E.~Witten, ``A Strong coupling test of S duality,'' Nucl.\
Phys.\ B {\bf 431}, 3 (1994) [arXiv:hep-th/9408074].}

\lref\ps{
J.~Polchinski and M.~J.~Strassler, ``The string dual of a confining
four-dimensional gauge theory,'' arXiv:hep-th/0003136.}

\lref\kkn{
V.~A.~Kazakov, I.~K.~Kostov and N.~A.~Nekrasov, ``D-particles, matrix
integrals and KP hierarchy,'' Nucl.\ Phys.\ B {\bf 557}, 413 (1999)
[arXiv:hep-th/9810035].}

\lref\kpw{
A.~Khavaev, K.~Pilch and N.~P.~Warner, ``New vacua of gauged $N = 8$
supergravity in five dimensions,'' Phys.\ Lett.\ B {\bf 487}, 14
(2000) [arXiv:hep-th/9812035].}

\lref\ks{
S.~Kachru and E.~Silverstein, ``4d conformal theories and strings on
orbifolds,'' Phys.\ Rev.\ Lett.\ {\bf 80}, 4855 (1998)
[arXiv:hep-th/9802183].  }

\lref\bj{
M.~Bershadsky and A.~Johansen, ``Large $N$ limit of orbifold field
theories,'' Nucl.\ Phys.\ B {\bf 536}, 141 (1998)
[arXiv:hep-th/9803249].}

\lref\lnv{
A.~E.~Lawrence, N.~Nekrasov and C.~Vafa, ``On conformal field theories
in four dimensions,'' Nucl.\ Phys.\ B {\bf 533}, 199 (1998)
[arXiv:hep-th/9803015].  }

\lref\bkv{
M.~Bershadsky, Z.~Kakushadze and C.~Vafa, ``String expansion as large
$N$ expansion of gauge theories,'' Nucl.\ Phys.\ B {\bf 523}, 59 (1998)
[arXiv:hep-th/9803076].}

\lref\vy{
G.~Veneziano and S.~Yankielowicz, ``An Effective Lagrangian For The
Pure $N=1$ Supersymmetric Yang-Mills Theory,'' Phys.\ Lett.\ B {\bf
113}, 231 (1982).  }

\lref\sinhv{
S.~Sinha and C.~Vafa, ``$SO$ and $Sp$ Chern-Simons at large $N$,''
arXiv:hep-th/0012136.  }

\lref\aahv{B.~Acharya, M.~Aganagic, K.~Hori and C.~Vafa,
``Orientifolds, mirror symmetry and superpotentials,''
arXiv:hep-th/0202208.}

\lref\dorey{N. Dorey, ``An Elliptic Superpotential for Softly
Broken $N=4$ Supersymmetric Yang-Mills,'' JHEP {\bf 9907}, 021 (1999)
[arXiv:hep-th/9906011].}

\lref\ls{
R.~G.~Leigh and M.~J.~Strassler, ``Exactly marginal operators and
duality in four-dimensional $N=1$ supersymmetric gauge theory,'' Nucl.\
Phys.\ B {\bf 447}, 95 (1995) [arXiv:hep-th/9503121].
}

\lref\klyt{
A.~Klemm, W.~Lerche, S.~Yankielowicz and S.~Theisen,
``Simple singularities and $N=2$ supersymmetric Yang-Mills theory,''
Phys.\ Lett.\ B {\bf 344}, 169 (1995)
[arXiv:hep-th/9411048]}

\lref\af{
P.~C.~Argyres and A.~E.~Faraggi, ``The vacuum structure and spectrum
of $N=2$ supersymmetric $SU(n)$ gauge theory,'' Phys.\ Rev.\ Lett.\
{\bf 74}, 3931 (1995) [arXiv:hep-th/9411057].}

\lref\mo{C.~Montonen and D.~I.~Olive,
``Magnetic Monopoles As Gauge Particles?,'' Phys.\ Lett.\ B {\bf 72},
117 (1977).  }

\lref\gvw{S.~Gukov, C.~Vafa and E.~Witten,
``CFT's from Calabi-Yau four-folds,'' Nucl.\ Phys.\ B {\bf 584}, 69
(2000) [Erratum-ibid.\ B {\bf 608}, 477 (2001)]
[arXiv:hep-th/9906070].  }

\lref\tv{
T.~R.~Taylor and C.~Vafa, ``RR flux on Calabi-Yau and partial
supersymmetry breaking,'' Phys.\ Lett.\ B {\bf 474}, 130 (2000)
[arXiv:hep-th/9912152].  }

\lref\mayr{
P.~Mayr, ``On supersymmetry breaking in string theory and its
realization in brane worlds,'' Nucl.\ Phys.\ B {\bf 593}, 99 (2001)
[arXiv:hep-th/0003198].  }

\lref\qfvy{A.~de la Macorra and G.G.~Ross, Nucl.\ Phys.\ B {\bf 404},
321 (1993)\semi C. P. Burgess, J.-P. Derendinger, F. Quevedo,
M. Quiros, ``On Gaugino Condensation with Field-Dependent Gauge
Couplings,'' Annals Phys. {\bf 250}, 193 (1996)
[arXiv:hep-th/9505171].}

\lref\div{
A. D'Adda, A.C. Davis, P. Di Vecchia and P. Salomonson, Nucl. Phys. B
{\bf 222}, 45 (1983).}

\lref\berva{
N. Berkovits and C. Vafa, ``$N=4$ Topological Strings,'' Nucl. Phys. B
{\bf 433} 123 (1995) [arXiv:hep-th/9407190].}

\lref\novikov{
V.~A.~Novikov, M.~A.~Shifman, A.~I.~Vainshtein and V.~I.~Zakharov,
``Instanton Effects In Supersymmetric Theories,'' Nucl.\ Phys.\ B {\bf
229}, 407 (1983).  }

\lref\dk{
N.~Dorey and S.~P.~Kumar, ``Softly-broken $N = 4$ supersymmetry in the
large-$N$ limit,'' JHEP {\bf 0002}, 006 (2000) [arXiv:hep-th/0001103].
}

\lref\ds{
N.~Dorey and A.~Sinkovics, ``$N = 1^*$ vacua, fuzzy spheres and
integrable systems,'' JHEP {\bf 0207}, 032 (2002)
[arXiv:hep-th/0205151].  }

\lref\ads{
I.~Affleck, M.~Dine and N.~Seiberg, ``Supersymmetry Breaking By
Instantons,'' Phys.\ Rev.\ Lett.\ {\bf 51}, 1026 (1983).  }

\lref\bele{
D. Berenstein, V. Jejjala and R. G. Leigh, ``Marginal and Relevant
Deformations of $N=4$ Field Theories and Non-Commutative Moduli Spaces
of Vacua,'' Nucl.Phys. B {\bf 589} 196 (2000) [arXiv:hep-th/0005087].}

\lref\gins{
P. Ginsparg, ``Matrix models of 2d gravity,'' Trieste Lectures (July,
1991), Gava et al., 1991 summer school in H.E.P. and Cosmo.
[arXiv:hep-th/9112013].}

\lref\kostovsix{
I. Kostov, ``Exact Solution of the Six-Vertex Model on a Random
 Lattice,'' Nucl.Phys. B {\bf 575} 513 (2000) [arXiv:hep-th/9911023].}

\lref\hoppe{
J.~Goldstone, unpublished; J.~Hoppe, ``Quantum theory of a massless
relativistic surface,'' MIT PhD thesis, 1982.}

\lref\superspace{S.J. Gates,
M.T. Grisaru, M. Ro\v{c}ek and W. Siegel, {\em Superspace}, Addison
Wesley 1983, hep-th/0108200.}

\lref\cov{M.T. Grisaru and D. Zanon, Nucl. Phys. B252 (1985) 578.}

\lref\dvi{R.~ Dijkgraaf, C.~ Vafa,
``Matrix Models, Topological Strings, and Supersymmetric Gauge
Theories,'' hep-th/0206255.}

\lref\dvii{R.~ Dijkgraaf, C.~ Vafa,
``On Geometry and Matrix Models,'' hep-th/0207106.}

\lref\dviii{R.~ Dijkgraaf, C.~ Vafa, ``A Perturbative Window
into Non-Perturbative Physics,'' hep-th/0208048.}

\lref\dhks{
N.~Dorey, T.~J.~Hollowood, S.~P.~Kumar and A.~Sinkovics, ``Massive
vacua of N = 1* theory and S-duality from matrix models,''
arXiv:hep-th/0209099; ``Exact superpotentials from matrix models,''
arXiv:hep-th/0209089.}

\lref\dhk{
N.~Dorey, T.~J.~Hollowood and S.~P.~Kumar,
``S-duality of the Leigh-Strassler deformation via matrix models,''
arXiv:hep-th/0210239.}

\Title{\vbox{\baselineskip11pt
\hbox{hep-th/0211017}
 \hbox{HUTP-02/A056}
 \hbox{ITFA-2002-47}
  \hbox{McGill/02-137}
\hbox{IFUM-734-FT}
  }}
  {\vbox{ \centerline{Perturbative
Computation of Glueball Superpotentials} \vskip 4pt
%\centerline{}{
}}
\centerline{
R. Dijkgraaf,$^1$
M.T. Grisaru,$^2$ C.S. Lam,$^2$ C. Vafa,$^3$ and D. Zanon$^4$}
%\medskip
%\medskip
%\medskip
\medskip
\vskip 8pt
\centerline{\it $^1$ Institute for Theoretical Physics \&
Korteweg-de Vries Institute for Mathematics,}
\centerline{\it University of Amsterdam, 1018 XE Amsterdam, The Netherlands}
\medskip
\centerline{\it $^2$ Physics Department, McGill University, }
\centerline{\it Montreal, QC Canada H3A 2T8}
\medskip
\centerline{\it $^3$
Jefferson Physical Laboratory, Harvard University,}
\centerline{\it Cambridge, MA 02138, USA}
\medskip
\centerline{\it $^4$ Dipartimento di Fisica Dell'Universit\'a
di Milano,}
\centerline{\it INFN, Sezione di Milano, Via Celoria 16, I-20133 Milano}
\medskip
%\medskip
\medskip
%\bigskip
\noindent Using ${\cal N}=1$ superspace techniques in four dimensions we show how to perturbatively compute the superpotential generated for the
glueball superfield upon integrating out massive charged fields.  The technique applies to arbitrary gauge groups and representations. Moreover we
show that for $U(N)$ gauge theories admitting a large $N$ expansion the computation dramatically simplifies and we prove the validity of the
recently proposed recipe for computation of this quantity in terms of planar diagrams of matrix integrals.

\smallskip
%\medskip

\Date{November, 2002}

%\draft

\newsec{Introduction}

The aim of this
 note is to use superspace techniques to compute
the glueball superpotential in perturbation theory in gauge theories
 with $\cN=1$ supersymmetry. In particular we will show that this computation
confirms the general prescription proposed in \dviii\ that these 
perturbative calculations can be done, for theories admitting
a large $N$ expansion,
 in a zero-dimensional reduction to a matrix
model. This conjecture has up to now only been proven in special cases where the corresponding theory could be engineered in string theory.

In this paper we basically make precise the field theory arguments sketched in \dviii\ by showing that the computations of the superspace Feynman
graphs for a chiral superfield in an external gauge field reduce to a zero-dimensional path-integral.  However, unlike in \dviii, we will not
limit ourselves to the case where the theory has a large $N$ expansion. 
 Instead we set up the computation for an arbitrary ${\cal N}=1$
supersymmetric gauge theory coupled to arbitrary matter. Moreover we show 
that for the case of $U(N)$ gauge theories admitting a large $N$
description this leads to the recipe advanced in \dviii .  The simplifications 
for other gauge groups and more general representations will appear
in a forthcoming paper
 \ref\another{R. Dijkgraaf, M.T. Grisaru, C.S. Lam, C. Vafa, and D. Zanon, to appear.}.
This leads to exact computation of F-terms for ${\cal N}=1$ supersymmetric
gauge theories in four dimensions for arbitrary gauge group and matter
content, upon extremization of the glueball superpotential.

The organization of this note is as follows: In section 2 we briefly review the relevant superspace techniques \superspace.  In section 3 we set
up the general computation for a general gauge group coupled to matter.  In section 4 we note the simplification in the context of $U(N)$ gauge
theories coupled to matter admitting a large $N$ expansion and confirm the proposal of \dviii. We illustrate the proof by working out a concrete
example.

\newsec{Chiral Superspace Techniques}

We consider the action, in the presence of an external gauge field,
for a self-interacting massive chiral superfield $\Phi$ in some
representation of the gauge group
\eqn\qwe{
S(\Phi, \Phib)= \int d^4x d^4 \th\ \Phib e^V \Phi +\int d^4x d^2 \th
\ W(\Phi) + h.c.
}
where we used a gauge invariant pairing between $\Phi$ and $\Phib$,
and where $W(\Phi)$ is some gauge invariant superpotential.  We use
the conventions from \superspace\ and in particular, for the gauge
field strength we have
\eqn\cwa{
\cW_\a = i \Db^2 e^{-V} D_\a e^V,
}
and a corresponding expression for $\overline{\cW}_\ad$.  Here $ \Db^2
= \frac{1}{2} \Db^\ad \Db_\ad$ and similarly $D^2= \frac{1}{2}D ^\a
D_\a$.  The covariant derivatives can be chosen as $ D_\a = \partial /
\partial \theta^\a$, $\Db_\ad =\partial / \partial \bar{\theta}^\ad+i
\bar{\theta}^\ad \partial_{\a \ad}$. Other superspace facts are the relations
\eqn\dbb{
\Db^2 D^2 \Phi = \Box\Phi,\qquad
D^2 \Db^2 \Phib = \Box \Phib,
}
valid for (anti)chiral superfields.  Here we consider, for simplicity of presentation, a single gauge group, but the analysis can be easily
extended to products of gauge groups as well.

Perturbative calculations for this system are best carried out in the background field method and details can be found in \superspace.
Furthermore, one can develop {\em covariant supergraph} techniques \cov\ which extend and simplify ordinary supergraph calculations. Here we will
not need the full power of this method; we will rederive what is necessary for the computation of the glueball superpotential.

In the usual supergraph analysis for the present system one has to
deal with propagators $\<\Phi \Phi\>$, $\<\Phib \Phib\>$ and $\<\Phi
\Phib\>$.  This would be true if one is trying to compute an arbitrary
amplitude.  However, here we are interested in certain F-terms and as
we now explain (and can also be verified using covariant supergraph
techniques \cov ) only holomorphic propagators $\<\Phi \Phi\>$
contribute to these terms.

The argument for this is rather simple and well-known: terms appearing in the superpotential cannot depend on the coefficients of the anti-chiral
superpotential ${\overline W}(\Phib)$. This is so because we can promote each of those coefficients to an anti-chiral field, whose vev then gives
the coupling constants. Holomorphy tells us that these fields cannot appear in an integral over chiral superspace. Since we are interested in
computing the superpotential for a chiral glueball superfield, we thus know that the antichiral coefficients do not contribute to this amplitude.

We use this fact to choose a simple form for ${\overline W}(\Phib)$
which is not necessarily the complex conjugate of
$W(\Phi)$. (Holomorphy allows us to treat them independently.)
Moreover, we assume that the fields are such that we can give them a
bare mass.  For simplicity, say, $\Phi$ is taken to transform in a
real representation so we can have a gauge invariant (necessarily
chiral) mass term ${m \over 2} \F^2$.  For $\overline W$ we now take
the simple form
\eqn\wphi{
{\overline W}(\Phib)= \hf \mbar\Phib^2.
}
(We could have also restricted to ${\overline W}=0$ and we will
comment on that possibility below). With this choice, the $\Phib$
action is quadratic and the antichiral superfield $\Phib$ can be
integrated out, leaving us to deal only with the chiral superfield
$\F$.

It is economical, in considering the kinetic term $\Phib e^V \Phi$ to
define the {\em covariantly antichiral} superfield
\eqn\phitilde{
\tilde{\Phi} =
\Phib e^V = e^{-V}\Phib,}
(the last identity holds for real representations). In this so-called
{\em gauge chiral representation} \superspace\ this field is trivially
annihilated by the {\em covariant} derivative $\Del_\a = e^{-V} D_\a
e^V$.  The corresponding covariantly chiral superfield is identical to
the ordinary chiral superfield and it is annihilated by the ordinary
derivative $\Delb_{\ad} \equiv \Db_{\ad}$. We have identities such as
$\Delb^2 \Del^2 \Phi = \Box_+ \Phi $ where, by straightforward
algebra,
\eqn\boxphi{
\Box_+ \Phi =\left[\Box_{\rm cov} -i \cW^\a
\Del_\a -\frac{i}{2}(\Del^\a \cW_\a)\right]\Phi.
}
There is a corresponding relation $\Del^2 \Delb^2 \tilde{\Phi }= \Box_- \tilde{\Phi}$ with $\Del^2 \Box_+ = \Box_- \Delb^2$. In the expression
above $\Box_{\rm cov}= {1 \over 2}\Del^{\a \ad} \Del_{\a \ad}$ and $\Del_{\a \ad} = -i\{ \Del_\a , \Delb_\ad\}$.

After these preliminaries we consider integrating out the antichiral
field from the action
\eqn\sphiphi{
S(\Phi, \Phib)= \int d^4x d^4 \th \ \tilde{\Phi} \Phi +\int
d^4x d^2 \th \ W(\Phi) + \int d^4x d^2 \bar{\th}\ {\mbar\over 2} \tilde{
\Phi}^2.}
Note that since the antichiral mass term must be a gauge singlet we were allowed to replace the quadratic term $\Phib^2$ by $\tilde{\Phi}^2$ ({\it
e.g.}, consider the case of the adjoint representation where $\Phib$ is an adjoint matrix, $\tilde{\Phi} = e^{-V}\Phib e^V$ and the $V$-dependence
drops out of $\Tr\,\tilde{\F}^2$.)  In standard fashion we wish to complete the square, but first we rewrite
\eqn\dxxx{
\int d^4x d^2 \bar{\th}\ \tilde{\Phi}^2 = \int d^4x d^4 \th \
\tilde{\Phi} \frac{1}{\Box_+} \Delb^2 \tilde{\Phi},
}
which can be checked by the replacement $\int d^2 \th \rightarrow D^2=\Del^2$ when acting on a gauge singlet.

We now complete the square as
\eqn\sqrr{
\int d^4x d^4 \theta \left\{\frac{\mbar}{2} \left[ \tilde{\Phi} + \frac{1}{\mbar} \Del^2 \Phi \right] \frac{1}{\Box_+}\Delb^2 \left[ \tilde{\Phi}
+ \frac{1}{\mbar}\Del^2 \Phi \right] - \frac{1}{2\mbar} \Del^2 \Phi \frac{1}{\Box_+}\Delb^2 \Del^2 \Phi\right\}.
}
We integrate by parts the $\Del^2$ and by the identities written above
get rid of the $1/\Box_+$. The antichiral superfield can be integrated
out now. In the last term we replace $\int d^2 \bar{\th}$ by $\Delb^2$
ending up with the action
\eqn\action{
S(\Phi) = \int d^4x d^2 \th \ \frac{-1}{2\mbar} \Phi \left[
\Box_{\rm cov} -i \cW^\a \Del_\a -\frac{i}{2}(\Del^\a \cW_\a)
\right] \Phi + { W}(\Phi).
}
%

%The Feynman rules follow: the propagator is
%$$
%< \Phi (x, \th) \Phi (x' \th')> = \frac{\mbar}{ [\Box_{\rm cov} -i W^\a
%\Del_a -\frac{i}{2}(\Del^\a W_\a) + m \mbar]} \d^4(x-x')\d^2(\th
%-\th')
%$$
%and (self-interactioin) vertices are obtained from the tree-level
%superpotential. Interactions with the background gauge field are
%obtained by separating from the propagator and expanding terms which
%depend on it.
For the purpose of determining the effective potential of a {\em
constant} singlet gaugino condensate $S \sim \cW^\a \cW_\a$ several
simplifications are now possible:

(a) We can assume that $\cW^\a$ is {\em covariantly} constant,
\eqn\covcnst{
\Del_{\a \ad} \cW^\b = \partial_{\a \ad}\cW^\b - i [ \Gamma_{\a \ad},
\cW^\b ] =0.
}
This implies, in particular, that $\cW$ commutes with $\Box_{cov}$ and allows us to move it freely from one line to another in a loop, at least
as far as the spacetime part is concerned.

(b) We can drop the term $\Del^\a \cW_\a$ which will never enter in a
relevant F-term.

(c) We can go to a new {\em gauge antichiral} basis by rewriting
\eqn\boxcov{
\Phi [\Box_{\rm cov} -i \cW^\a e^{-V}D_\a e^V ] \Phi = \Phi' [\Box'_{\rm
cov} -i \cW'^\a D_\a ] \Phi',
}
where $\Phi' = e^V \Phi$ and $\cW' = e^V \cW e^{-V}$. The functional integral for $\Phi'$ is the same as for $\Phi$ (it is a local field
redefinition) while in the gaugino condensate $\Tr\,\cW^2 =\Tr\, \cW'^2$.

(d) The connection terms in $\Box_{\rm cov}$ can be treated perturbatively and in general would serve to covariantize derivative terms in the
effective action, of which there are none in the present situation. (This is not entirely true: spinor derivatives of the space-time connection
can lead to field strengths but a more detailed covariant supergraph analysis shows that this is not the case here.) Therefore the connections can
be dropped from $\Box_{\rm cov}$ and we can replace it with the ordinary d'Alembertian $\Box$.

Summarizing, we can conclude that for our purpose of computing F-terms
in an external gauge field the relevant action can be written as
\eqn\lagr{
\int d^4x d^2\theta\ \left( {-1\over 2\mbar} \F (\Box -i \cW^\a D_\a)\F
+ W_{\rm tree}(\F)\right).  }

\newsec{The Perturbative Computation}

The action \lagr\ derived above will be our starting point for the
perturbative computation of the effective superpotential in the $\cW$
background. We will write the tree-level superpotential as
\eqn\wtree{
W_{\rm
tree}(\F) = {m\over 2}\F^2 + \hbox{\it interactions}.
}
It is convenient to include the $m\F^2$ part of the superpotential in
the propagators, as usual.

\subsec{Localization and $\mbar$-independence}

One important aspect of the above action is the fact that the
superfield $\cW^\a$ is correlated with the spinor derivative $D_\a$,
and this confirms the approach used in \dviii.  In \dviii\ it was also
suggested that localization of the path integral to constant modes
arises by asking which configurations can contribute to a term that
can be written as an integral over chiral superspace. One used the
argument familiar from topological field theory that derivatives can
be written as commutators with the antichiral supercharge
$\overline{Q}_{\ad}$ and therefore do not contribute in a $\int\!
d^2\th$ term.  Let us see if this latter conclusion is valid in the
present setup.

As we discussed before, the above path-integral should be $\mbar$
independent.  We will verify this directly below, in perturbation
theory.  Since the answer does not depend on $\mbar$, one could even
take the limit $\mbar \to 0$, {\it i.e.}\ set the antichiral
superpotential ${\overline W}(\Phib) = 0$. Because $1/\mbar$
multiplies the kinetic term in the Lagrangian \lagr , in this case the
path-integral over the chiral field $\F$ will localize to the
solutions of
\eqn\constraint{
(\Box -i \cW^\a D_\a)\F=0.
}
In the absence of a $\cW^\a$ background this would imply (in a Euclidean path-integral) a localization to harmonic and therefore constant $\F$.
However, for us it is crucial to have a non-vanishing $\cW^\a$ background turned on. So we see that it is not quite correct to say that the
localization to constant modes takes place when this background is turned on. There will be non-trivial $x$ (or $p$) dependence, although the
solution space to the constraint \constraint\ is still finite dimensional.
%In fact, solving the above equation leads
%to momentum contributions localized near
%$$
%(p^2)^2 \sim  F_+ \wedge F_+ .
%$$
%(This can be best seen by working in components. EXPLAIN. ALSO DEFINE F)
One can then localize the path-integral near these configurations and compute the partition function.  However, instead, it turns out to be easier
to directly do the Feynman graphs of this theory.

Let us first show by direct computation why the $\mbar$ dependence
cancels out.
%  This cancellation comes from the momentum integrals, as well as the
%$\cW^\a D_a$ insertions.
Consider a Feynman graph made of chiral superfields with a total of
$\ell$ loops.  We will have one 4-momentum integral $\int \! d^4p$ for
each loop.  Moreover, we can also go to a momentum representation in
the $\theta^\a$-directions and write
\eqn\dpi{ D_\a = \partial/ \partial \theta^\a=-i\pi_\a }
leading to $2\ell$ grassmann momentum integration variables.

The bosonic part of the propagator is $\mbar/(p^2+m\mbar)$ and we can remove its $\mbar$ dependence by rescaling $p^2 \to \mbar p^2$. Then the
bosonic $d^4p$ momentum integrals will rescale with a factor of ${\mbar}^{2\ell}$. On the other hand, to absorb the fermionic momentum
integrations $\int \! d^2\pi$ we need $2\ell$ interaction factors $\cW^\a \pi_\a$ from the action and each comes with a factor of $1/\mbar$ from
the action or, equivalently, since we expand the propagator
\eqn\propa{
{\mbar \over p^2 + m\mbar + \cW^\a \pi_\a}
}
in an external background.  Thus we obtain a factor of $1/\mbar^{2\ell}$ from the fermionic momentum integration and we see that the factors of
$\mbar$ cancel between bosonic and fermionic momentum integrations. Below we will use this freedom to scale $\mbar$ and set it to $\mbar =1$. As
an aside we remark that in computations of F-terms the parameter $\mbar$ plays very much the same role as the the parameter $\beta$ (the inverse
temperature) does in the usual heat kernel arguments that compute the anomaly or Witten index (more precisely, $\mbar \sim 1/\beta$). One can send
$\mbar$ either to zero or infinity and thus relate different expansions.

Keeping track of the grassmann integrals also leads to another
important restriction valid for F-terms: a diagram with $\ell$ loops
will contribute precisely a factor of $\cW ^{2\ell}$ (with various
possible gauge and spinor index contractions) to the effective
superpotential, irrespective of the gauge group and the matter field
representations.

\subsec{Loop momentum integrals}

Our task now is to do the actual momentum integrals and compute the
Feynman diagrams. Here we take the hint from how the computation of
the glueball superpotential is done in the string context \bcov .
There, in particular, one uses worldsheet moduli, describing the
inequivalent conformal structures on the worldsheet. For the string
theory argument it is important that one integrates first over the
spacetime momenta in the loops and only then over the geometric
moduli. The integral over the four-dimensional loop momenta gives a
factor
\eqn\detomega{
{1 \over \det(\Im \Omega)^2},
}
with $\Omega_{ab}$ the period matrix of the Riemann surface that
represents the string worldsheet.  This measure factor gets cancelled
when one does the fermionic momentum integration, as is most easily
seen in the Berkovits type formalism applied to this problem in
\berva. Thus for these terms in the effective action there is a
drastic simplification in the integrand of the string measure on the
worldsheet moduli space --- the complete dependence on the
four-dimensional kinematics disappears. It is this simplification that
allows one to reduce the computation to that for a zero-dimensional
matrix integral and recover the prescription of \dviii.

In the corresponding field theory setup this suggests that in order to
see the simplification one should represent the Feynman amplitude in
terms of Schwinger time variables, which are the field theory limits
of worldsheet moduli --- the Schwinger parameter $s$ associated to a
propagator can be thought of as the length of an edge in the Feynman
graph.

For any given Feynman graph we have a number of Schwinger time parameters $s_i$ where $i$ runs over the edges of the Feynman diagram. We can think
of the $s_i$ as parametrizing the moduli space of the diagram, {\it i.e.}\ the metric we can put on the graph considered as a one-dimensional
(singular) space. They are the QFT analogues of the string worldsheet moduli \ref\pol{ J.~Polchinski, ``Evaluation Of The One Loop String Path
Integral,'' Commun.\ Math.\ Phys.\ {\bf 104}, 37 (1986).}, see also {\it e.g.}\ \ref\lam{C.S. Lam, ``Multiloop String-Like Formulas for QED,''
Phys.~Ref.~ D 48 (1993) 873 [arXiv:hep-ph/9212296].}.  In momentum space the propagators are represented as
\eqn\schwinger{
\int_0^\infty \!\! ds_i \ \exp\left[-s_i\left(p_i^2+\cW^\a\pi_{i\a} +
m\right)\right], }
where we have included the $m$-dependence in the propagator and have
put $\mbar=1$.  Now let $p_a$ be the loop momentum for the $a$-th
loop, and let
\eqn\pLp{
p_i= \sum_a L_{ia}\, p_a }
denote the total momentum flowing through the $i$-th propagator, where
we can normalize the matrix entries $L_{ia}$ to $0,\pm 1$ .  In this
notation the Schwinger action for the $i$-th edge is
\eqn\expsL{
\exp\Bigl[ -s_i \Big(\sum_a L_{ia} p_a\Bigr)^2\Bigr].
}
Let us also introduce the matrix $M_{ab}(s)$ given by
\eqn\mmm{
M_{ab}(s)= \sum_i s_i L_{ia} L_{ib}.  }
This matrix naturally appears if we write the total first-quantized
action in terms of the loop momenta $p_a$ as
\eqn\exppMp{
\exp\Bigl[ - \sum_{a,b} p_a M_{ab}(s) p_b \Bigr].
}
We can perform the gaussian integral over bosonic momenta in all loops
and obtain
\eqn\boson{
Z_{\rm boson} =\int \prod_{a=1}^\ell {d^4p_a \over (2\pi)^4}\
\exp\Bigl[ - \sum_{a,b} p_a M_{ab}(s) p_b \Bigr] =
{1\over (4 \pi)^{2\ell}} {1\over \(\det M(s)\)^{2}}.
}
Thus, the factor $(\det M(s))^{-2}$ gives the measure on the field
theoretic Schwinger moduli space. Note that the matrix $M_{ab}$ can be
identified as the field theory limit of the period matrix
$\Im\Omega_{ab}$ in the worldsheet theory, and the above measure is
the QFT analogue of \detomega.

For the fermionic momenta we can do something very similar. Let
$\pi_{a}^{\a}$, ($a=1,\ldots,\ell$, $\a=1,2$) denote the fermionic
spinor momentum running around the $a$-th loop. The total
$\pi$-momentum flowing through the $i$-th propagator is then given
again by
\eqn\piLpi{
\pi_{i\a}=\sum_a L_{ia}\pi_{a\a}.
}
We now have to perform the grassmann integrals
\eqn\ferm{ \int \prod_{a=1}^\ell d^2\pi_a\
\exp \Bigl[-\sum_i s_i \Bigl( \sum_a \cW_i^\a L_{ia}\pi_{a\a}\Bigr)\Bigr].
%\times \hbox{\it group theory factor}
}
This expression gets multiplied with the relevant group theory
factor. Here $\cW_i^\a$ means the restriction of $\cW^\a$ to the
representation dictated by the $i$-th edge and this should be viewed
as an operator on the $i$-th representation vector space.  So a more
explicit notation would be
\eqn\wiaa{
\cW_i^\a = \sum_A \cW^\a_A T^A_i,}
where $A$ is a Lie algebra index and $T_i^A$ denotes the corresponding Lie algebra generator in the representation propagating through the $i$-th
edge. The evaluations of the term \ferm, {\it i.e.}\ the integration over the fermionic loop momenta $\pi_a$, will very much depend on the
representation structure of the edges and various group theory factors, which can in principle be analyzed. This will be presented elsewhere
\another. In the next section we see a dramatic simplification for the $U(N)$ case with adjoint representation (or fundamental representations)
which verifies the matrix integral proposal of \dviii .

\newsec{Proof of the Matrix Integral Reduction for $U(N)$}

For the case of a $U(N)$ gauge theory interacting with a matter field
in the adjoint representation (which can be easily generalized to quiver
type theories involving product of $U(N_i)$)
 we can use the 't
Hooft double line notation to keep track of the gauge index
structure. The 't Hooft diagrams at $\ell$ loops have at most
$h=\ell+1$ holes (including the outer boundary).

Suppose now we are interested in computing the superpotential for the
{\it traceless} glueball field
\eqn\SS{
S={1\over 32 \pi^2} \Tr_{SU(N)} {\cW^\a \cW_\a }.
}
So we explicitly exclude higher order traces of the form $\Tr\,\cW^n$ with $n>2$. Since $S$ is given by the quadratic form $\Tr\, \cW^2$, this
means that in the 't Hooft double line notation we want for each index loop or hole exactly two insertions of $\cW$. Thus we need at least $\ell$
index loops, since, as we discussed in the previous section, at $\ell$ loops in field theory we will bring down $2\ell$ factors of $\cW$ in the
Schwinger representation of the Feynman diagram. But this can only happen if the diagram is planar, as was already noted in \dviii . Planarity is
therefore enforced by the fermionic integrals, independent of the precise 
measure induced by the loop momenta.

Note that there is necessarily one index loop where $\cW$ is {\it not} inserted.
 We have $h=\ell+1$ index loops and we have to choose in which
index loop we will not insert any $\cW$'s.  This can clearly be done in $h$
 ways and this gives a combinatorial factor $h$ multiplying the
diagram. We also get a factor of $N$ for the single index loop which has 
no fields attached to it.

Once all this is done we can associate each $\cW_i$ as being part of a
``$\cW$-loop.''  Such an index loop is in general formed by
multiplying and finally tracing a series of $\cW$ insertions
\eqn\traces{
\Tr\left(\cW_{i_1}\cdots \cW_{i_n}\right).
}
But in the present case, where we only consider the dependence on the
bilinear $S= {1\over 32\pi^2}\Tr\,\cW^2$, such a loop only consists of
two $\cW$'s. After we have done the $\cW$ insertions, two per index
loop with fixed and opposite spinor indices, the complete group index
structure is entirely captured by just multiplying each loop by a
factor $16 \pi^2 S$.

Note that quite generally, also for the generating function of traces of higher powers of $\cW$, we can have for every edge at most two $\cW$'s
inserted. To see this for the $i$-th propagator, we can use the fermionic loop momentum $\pi_i$ as one of the fermionic integration variables.
Doing the integral $\int\!d^2\pi_i$ will then bring down at most two $\cW$ insertions.

Furthermore we should recall that the fields are in the adjoint representation of $U(N)$ and  therefore the action of $\cW$ is through
commutators. That is, in the Schwinger action we have the term
\eqn\commutator{
\exp\Bigl( -s_i \big[W_i^\a,-\big]  \pi_{i\a}\Bigr).
}
Because of this commutator the insertions at the two double lines come with opposite signs. As we have stated already, for a planar diagram we can
identify the index loops with the momentum loops (except for one, say the outer loop). Every index loop has a canonically induced orientation
coming from the orientation of the plane, and we use the same orientation to assign the direction of the loop momenta. In this way the two lines
that make up the double line propagator have opposite orientation. Note that because of the commutator the $\cW$ insertions on the left or right
index line have signs that are correlated with the index loop orientations.

We can now use this fact, correlating the index loops and momentum loops, to keep track of these group index contractions in a way that is strictly
analogous to the fermionic loop momenta $\pi^\a_a$. We will introduce a set of auxiliary grassmannian variables $\cW^\a_a$ for each loop and write
\eqn\cwais{
\cW^\a_i= \sum_a L_{ia} \cW^\a_a.
}
The cases $L_{ia}=\pm 1$ correspond to the left and right commutator action
respectively.  Contracting the group indices is now equivalent to
performing the fermionic integration over the auxiliary variables
$\cW^\a_a$. It gives automatically precisely two $\cW$ insertions per
index loop, and it also assigns the right signs.

This trick allows us to represent the fermionic contribution \ferm\ to
the Feynman diagram as an integral over the grassmannian loop momenta
$\pi^\a_a$ and ``index momenta'' $\cW^\a_a$
\eqn\fff{\eqalign{
Z_{\rm fermion} = & Nh (16\pi^2 S)^{h-1} \int \prod_{a}d^2\pi_{a}d^2\cW_a
\ \exp\Bigl[-\sum_i s_i \Bigl( \sum_{a,b} \cW_a^\a L_{ia} L_{ib}\pi_{b\a}
\Bigr)\Bigr] \cr
= & Nh (16 \pi^2 S)^{h-1} \int \prod_{a}d^2\pi_{a}d^2\cW_a\
\exp\Bigl[-\sum_{a,b} \cW_a^\a M_{ab}(s)\pi_{b\a}\Bigr] \cr
= &  Nh S^{h-1} (4 \pi)^{2h-2} \(\det M(s)\)^2\cr }}
This should finally be multiplied with the corresponding bosonic
measure factor \boson. Using the planarity relation $\ell=h-1$ that
contribution can be written as
\eqn\zbozon{
Z_{\rm boson}= {1\over (4 \pi)^{2h-2}} {1\over (\det M(s) )^{2}}.
}
We see that all $s$ dependence cancels between bosons and fermions
giving for a diagram with $h$ holes the simple factor
\eqn\zfermion{ Z_{\rm boson} \cdot Z_{\rm fermion} = Nh S^{h-1}.
}

This cancellation is the field theory reflection of the fact that we are computing an amplitude that within string theory reduces to a topological
string amplitude. Note that the straightforward expansion in $\cW$'s leads to contributions of many different Feynman diagrams
--- we can distribute the $\cW$'s over the different propagators and
within each propagator over the two double lines.  The kinematic weight for each of the individual diagrams can be very complicated, which one
discovers immediately by computing a specific diagram. Only for the sum of all these diagrams do the fermionic and bosonic measures cancel. Of
course, this sum of diagrams is realized by one string worldsheet, and that is why the cancellation argument is so elegant and direct within the
string setup. In the field theory limit we localize to various corners of the moduli space and the simplification is less explicit.

With the cancellation of the bosonic and fermionic measures we are
thus left with the contribution of the holomorphic mass $m$. For each
Schwinger time $s_i$ we have the simple integral
\eqn\swing{
\int ds_i \ e^{-s_i m}={1 \over m}.
}
This reproduces the propagator of the zero-dimensional action ${m\over 2}\Tr\,\F^2$. Including also the coupling constants from the vertices we
have thus obtained the same amplitude as the planar graphs of a matrix model with $h$ boundaries multiplied by $NhS^{h-1}$ as was proposed in
\dviii. Hence, we have derived the following formula for the perturbative contribution of the effective glueball superpotential
\eqn\weff{
W_{\rm pert}(S)=N{\partial \cF_0 \over \partial S},
}
with
\eqn\cfo{
\cF_0(S)= \sum_h \cF_{0,h}S^h,
}
where $\cF_{0,h}$ is the matrix model planar amplitude with $h$ holes.
Note that the full answer for the effective superpotential also
includes the Veneziano-Yankielowicz term $N S \log(S/\Lambda^3)$ \vy,
which in the matrix model comes from the volume of the gauge group
\ov.

We could also be asking about contributions of the form
\eqn\taus{
{1\over 2}\int d^4x d^2\theta\  \tau(S) \Tr\,\cW^\a \, \Tr\,\cW_\a .
}
This kind of amplitude computes the $U(1)$ coupling constants upon
gaugino condensation.  In this case, as noted in \dviii, the 't Hooft
index structure is such that on two of the index loops we have to put
a single $\cW$ while the rest get two each.  There are $h(h-1)/2$ ways
to do that.  Once this is done the rest of the computation is
identical to the one above.  Thus in particular we find
\eqn\tauss{
\tau (S)={\partial^2 \cF_0 \over \partial S^2},
}
as was proposed in \dviii.

Note that the above computation will also give rise to contributions
involving $\cW$ with a different index contraction structure. In
particular, an $\ell$-loop gauge theory Feynman diagram, when written
in double line notation, can correspond to a surface with $g$ handles
and $h$ holes as long as
\eqn\ggg{
2g-2+h=\ell-1.
}
with the total of $2\ell$ insertions of $\cW$ distributed among the
$h$ holes. For such a general amplitude the computation will be similar
to the above computation done for planar diagrams. A more natural role for the
non-planar matrix model diagrams come from gravitational couplings to
RR-field strength which capture the lack of anti-commutativity in the
$\theta$-directions for the gauge theory \ref\oogv{H. Ooguri and
C. Vafa, work in progress.}.

\subsec{An example}

\ifig\stopsign{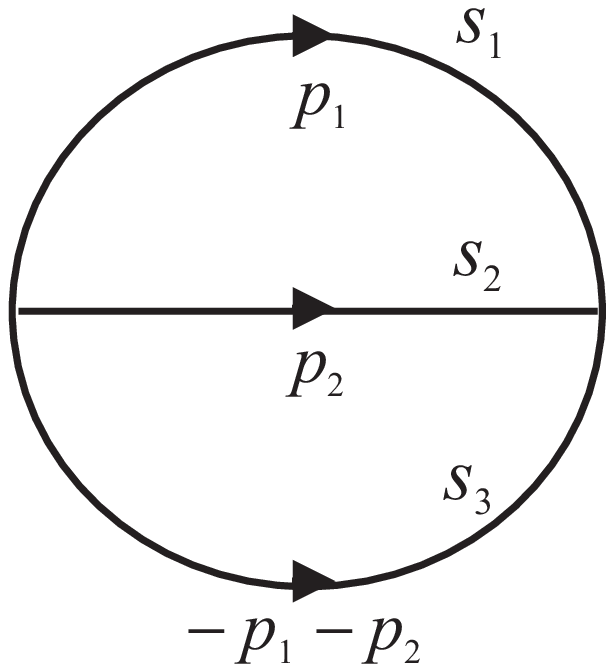}{30}{
A two-loop diagram: $s_i$ are the Schwinger parameters, the loop
momenta flowing through the propagators are indicated.}

Just to illustrate how the computation works in a specific case we present a simple example here. Let us consider the model with a single adjoint
chiral multiplet and a cubic superpotential
\eqn\phicube{
W(\F) = \Tr\left(\hf m \F^2 + {1\over 6}g \F^3\right).
}
In this model we will now compute what is essentially the first
non-trivial diagram --- the ``stop sign'' two-loop diagram depicted in
\stopsign. In this diagram there are three propagators with Schwinger lengths
$s_1,s_2,s_3$. There are two independent loop momenta $p_1,p_2$. We
will pick a basis such that the bosonic momenta flowing through the
three edges is $p_1,p_2,-p_1-p_2$ respectively, with similar
expressions for the fermionic loop momenta $\pi_i$. So in this case
our matrices $L_{ia}$ and $M_{ab}$ are given by
\eqn\Lmatrix{
L_{ia} = \pmatrix{1 & 0 \cr
          0 & 1 \cr
          -1 & -1 \cr }
}
and
\eqn\Mmatrix{
M_{ab} = \pmatrix{s_1+s_3 & s_3 \cr
          s_3 & s_2+s_3\cr}.
}

\ifig\ins{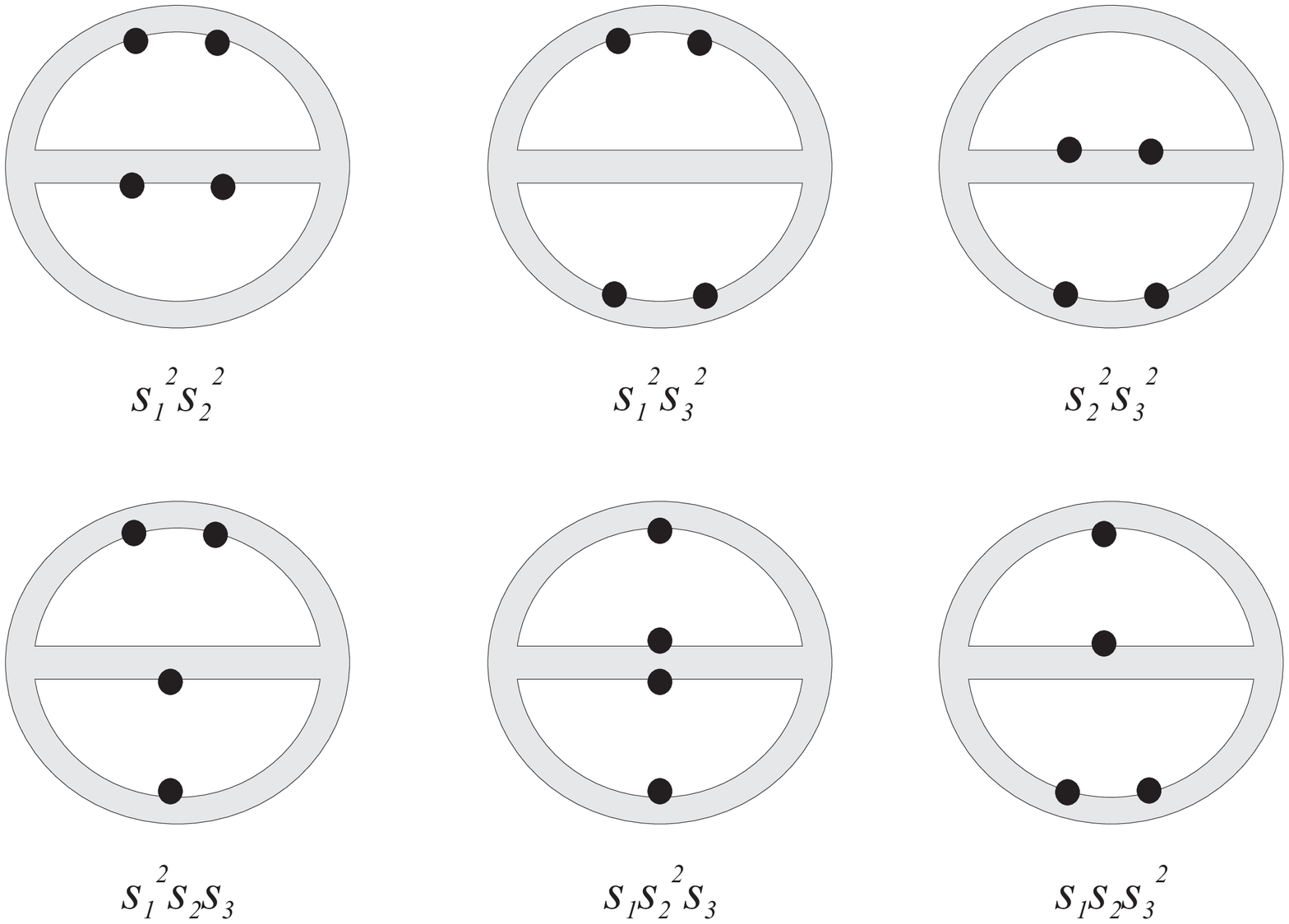}{100}{The diagrams with $\cW$ insertions
(indicated by $\bullet$'s) and the monomials in the Schwinger
parameters they compute. We fixed the outer index loop to be free.}

\noindent The integral over the bosonic loop momenta takes the form
\eqn\looppp{
\int {d^4p_1\over (2\pi)^4} {d^4p_2 \over (2\pi)^4}
\ \exp\left[-s_1p_1^2-s_2 p_2^2 -s_3(p_1+p_2)^2\right],
}
and this gives the measure factor
\eqn\mom{
{1\over (4\pi)^4} {1\over (\det M(s))^2}={1\over (4\pi)^4} {1\over
(s_1s_2 + s_2s_3 + s_3s_1)^2} }
To find the $\cW$ dependence we have to evaluate now the fermionic
loop momentum integral. Instead of using the representation in our
proof, let us compute this ``by hand'' by summing over all possible
$\cW$ insertions. An insertion on the $i$-th propagator gives a factor
$s_i$. There are at most two insertions on one edge.

If we fix one of the index loops to be free of insertions, say the
outer loop, then there is precisely one diagram for each monomial in
the $s_i$'s, see \ins.  Note that every factor $s_i^2$ comes with an
extra factor of $\hf$ because we expand the exponential to second
order.  In this way we find the weight
\eqn\nextol{
{3\over 4} \(\Tr\,\cW^2\)^2 \( s_1^2s_2^2 + 2 s_1^2 s_2 s_3 + \hbox{\it
permutations}\),
}
which can be written as
\eqn\thlas{
3S^2\cdot (4\pi)^4 \cdot (s_1s_2 + s_2s_3 + s_3s_1)^2
}
Note the factor $h=3$ that comes from choosing one of the three index
loops on which one does not insert $\cW$'s. This fermionic
contribution cancels exactly the momentum integral \mom, as promised.

\bigskip
\centerline{\bf Acknowledgements}

We would like to thank J. de Boer, S. Gukov, P. van Nieuwenhuizen
and H. Ooguri for
valuable discussions. R.D. wishes to thank Harvard University for kind
hospitality during this work.

The research of R.D.~is partially supported by FOM and the CMPA grant of the University of Amsterdam; M.T.G~is supported by NSF grant PHY-0070475;
C.S.L.~is supported in part by NSERC, Canada and Fonds de Recherche sur la Nature et les Technologies du Quebec; C.V.~is partially supported by
NSF grants PHY-9802709 and DMS-0074329; and D.Z.~is partially supported by INFN, MURST, and the European Commission RTN program
HPRN-CT-2000-00113.

\listrefs

\bye